# An Analytic Model to Determine the Interstitial-Solute Energetics and Underlying Mechanism in Refractory High-Entropy Alloys

**Authors**: Qianxi Zhu[1], Wang Gao[1*], and Qing Jiang[1]

[1] Key Laboratory of Automobile Materials, Ministry of Education, Department of Materials Science and Engineering, Jilin University, Changchun 130022, China.

*Corresponding author(s). E-mail(s): wgao@jlu.edu.cn.

**Abstract:** The solution and diffusion of interstitial non-metallic solutes (INSs) like H, He, O, C, N, P, and S is common in refractory high-entropy alloys (RHEAs) and essentially controls the RHEAs properties. However, the disorder local chemical environments of RHEAs hinder the quantitative prediction of the stability and diffusivity of INSs and the understanding of the underlying mechanism. Based on the tight-binding models, we propose an analytic model for determining the stability and diffusivity of INSs in RHEAs, by approximating the bonding length between INSs and their neighbors with the atomic radius of the neighbors in elemental states. This predictive model identifies that the energetics of INSs depends linearly on the d-band width of their neighbors, with the slope determined by the valence of INSs. Our scheme provides an electronic-level understanding of INSs in RHEAs and explains key experimental observations, which can serve as an effective tool for designing advanced RHEAs.

    The solution and diffusion of interstitial non-metallic solutes (INSs) have a significant impact on the structural stability, mechanical properties and functional characteristics of metals[1–6]. The non-metallic atoms such as H, He, O, C, N, P, and S atoms can easily enter into lattice interstitial sites in the preparation, processing, and practical use of metals[7–9]. Notably, the dissolution of interstitial H in metals is crucial to the H-embrittlement and solid-state H storage that have been key factors limiting the performance of nuclear materials and the application of hydrogen energy[10–12]. Recently, refractory high-entropy alloys (RHEAs) with body-centered cubic (BCC) structures are found to exhibit excellent high-temperature stability, strong resistance to radiation damage, high-performance H storage and so on[13–19]. It is thus particularly important to identify the determinants of the stability and diffusivity of interstitial H in RHEAs. However, the complex elemental composition and disordered local chemical environments of RHEAs prohibit understanding the solution and diffusion of INSs in general, and interstitial H in particular. Therefore, it is of great significance to reveal the fundamental coupling mechanism of INSs and metal atoms in RHEAs, which enables to build a quantitative model for predicting the stability and diffusivity of INSs that facilitating the development of high-performance RHEAs.

    In the BCC pure metals, the periodic arrangement of the lattice atoms creates equivalent interstitial sites, with interstitial H preferring to occupy tetrahedral interstitial sites (TISs)

compared to octahedral interstitial sites (OISs), and interstitial H diffusion primarily occurs between TISs along the <110> direction[20–27]. However, in RHEAs, the solution and diffusion of interstitial H exhibit significant differences even for TISs or OISs, due to the variations of the local chemical environment[28]. Previous studies primarily adopted the interstitial volume and the valence-electron concentration (*VEC*) of the neighboring atoms to estimate the interstitial H solution energy of RHEAs[29–33]. For instance, Wang *et al.* divided the interactions of interstitial H with metal atoms into the elastic and chemical contributions, and further developed a model for the H solution energy of HEAs based on interstitial polyhedral volume[30]. However, this model involves multiple fitting parameters and exhibits limited accuracy. Similarly, Zhou *et al.* established a linear relationship between interstitial H solution energy and the reduced TIS volume in WTaVCr by dividing the TISs into different groups based on the H affinity of different elements[29]. Meanwhile, Nygård *et al.* proposed a correlation between the *VEC* of the neighboring atoms and the stability of TIS deuterium[31–33]. These studies provide important insights into the H solution and diffusion in RHEAs, but only with a qualitative assessment of energetics. Notably, Borges *et al.* used a local cluster expansion (CE) model to quantify the solution energies of interstitial O, C, and N atoms in RHEAs[34]. The CE model offers an effective approach for predicting interstitial solution energy; however, it involves a substantial number of system-specific parameters that generally lack clear physical significance, limiting its capacity for generalized predictions. Recently, machine learning models that integrate symbolic regression algorithms have been shown effective in predicting the H solution and diffusion of HEAs, which can provide analytical expressions[35,36]. However, these data-driven expressions are often complex and do not provide in-depth insights into the underlying mechanism. Clearly, it is urgently needed to build a quantitative prediction model from the easily accessible and physically meaningful properties for the solution and diffusion of INSs in RHEAs.

In this contribution, we derive an analytical expression of the interatomic matrix element $V_{is}^2$ for the coupling between INSs and constituent elements in RHEAs based on the tight-binding (TB) model[37–39], by determining the bonding distance between INSs and their neighboring metals using the intrinsic atomic radius of metal elements. $V_{is}^2$ enables the site-to-site prediction of the solution and diffusion of interstitial H, He, O, C, N, P, and S atoms in RHEAs, and rationalizes many experimental results, demonstrating good prediction accuracy and broad applicability. Notably, our model reveals that the interatomic coupling between INSs and alloying atoms in RHEAs is dominated by the d-band width of the alloy elements and the valence properties of INSs, providing a new physical framework for understanding the complex multi-element synergistic effects on the solution and diffusion of INSs in RHEAs. This model is thus of great helpful to accelerate the development of advanced RHEAs.

## Results

We systematically investigate the solution of INSs at the interstitial sites of various RHEAs, including NbTaWV, TaVCrW, NbTiZrW, NbMoTaW, NbTaTiW, NbMoTaTi, NbMoTaWV, NbTaTiWV, NbMoTaTiW and HfNbTaTiW. In our calculations, we observe that the interstitial H and He atoms tends to occupy the TISs in RHEAs due to their small atomic sizes (Fig. 1a), while interstitial O, C, N, P, and S atoms prefer to occupy the OISs in RHEAs to accommodate their larger atomic sizes (Fig. 1b), consistent with the previous studies[28,34,40]. Taking the TIS as the initial-state (IS) and final-state (FS), the H diffusion can proceed via the TIS-TIS (T-T) path along the <110> direction and TIS-OIS-TIS (T-O-T) path along the <100> direction (shown in Fig. 1c, d). In the BCC transition metals (TMs), the diffusion barrier ($E_b$) for interstitial H along the T-T path is lower than that of the T-O-T path[23,26]. However, the complex element distribution in RHEAs may change the competition between T-T and T-O-T path. Thus, both the T-T and T-O-T paths are investigated for H diffusion in RHEAs, along with the OIS-TIS-OIS (O-T-O) path for O diffusion.

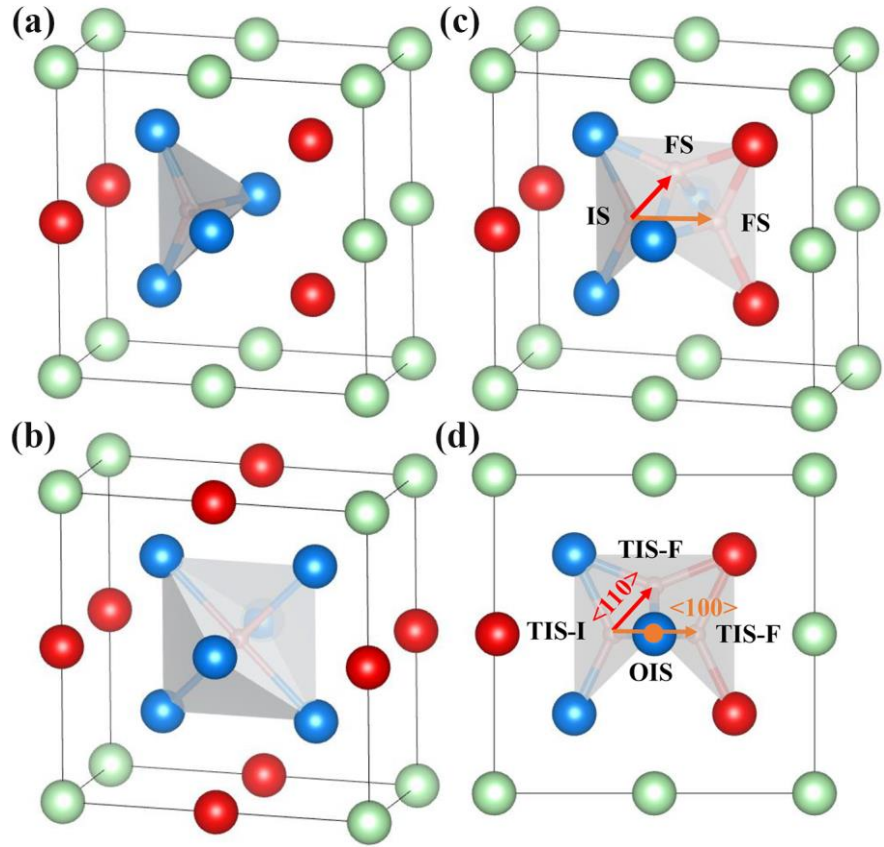

**Fig. 1 Schematic representation of the interstitial structures and the diffusion paths in RHEAs.** (**a-b**) The first nearest-neighboring (1nn) and second nearest-neighboring (2nn) atoms of interstitial non-metallic solutes (INSs) in (**a**) tetrahedral interstitial site (TIS) and (**b**) octahedral interstitial site (OIS) of the BCC RHEA structure. The blue spheres denote the 1nn atoms of INSs, while the red spheres represent the 2nn atoms. (**c-d**) The two diffusion pathways of interstitial H, TIS-TIS (T-T) and TIS-OIS-TIS (T-O-T) in BCC RHEAs. The difference in the constituent elements of RHEAs is not shown to better illustrate the difference in neighboring sites.

**Interstitial solution in RHEAs**

We first focus on the solution of interstitial H in TISs of RHEAs. The interstitial solution energy ($E_{sol}$) is an important parameter for determining the interactions between INSs and metal atoms. In RHEAs, differences in local chemical environments lead to significant variations in $E_{\text{sol,H}}$ between different TISs. Therefore, determining the local electronic structures of RHEAs is essential for understanding the coupling mechanism between interstitial H and alloy atoms, as well as for the site-to-site prediction of $E_{\text{sol,H}}$. We thus analyze the solution of interstitial H in RHEAs from the perspective of the intrinsic electronic structures. To better understand the interactions between interstitial H and surrounding metal atoms in RHEAs, we start from the solution of H atoms in TMs. The interactions of H atoms with TMs can be categorized into the coupling of the s-orbital of H with the s-band and d-band of TMs. The s-electrons of TMs exhibit a broad distribution and display characteristics similar to free electrons, while the d-electrons have a more confined distribution and display localized properties[37,41]. According to the TB model, the s-band of TMs is typically regarded as half-filled states, with the coupling between s-states of TMs approximated as a constant. In contrast, the coupling between d-states of TMs is significantly different from that of the s-states, playing a dominant role in determining the trend of bonding strength from one metal to another[37]. Here, we assume that the coupling between TMs and interstitial H follows a similar rule, with the trend dominated by the d-band of TMs. Therefore, we focus on the d-band contribution of TMs to the solution of interstitial H. The interatomic matrix element $V_{\text{is}}$ between the states of INSs like H and the d-states of TMs is determined with the following formula[37,38],

$$V_{\text{is}} = \eta_{\text{sd}} \frac{\hbar^2 r_{\text{is}} r_{\text{d}}^{3/2}}{ml^{7/2}} \tag{1}$$

where $r_{\text{is}}$ is the radius of interstitial-solute states, $r_{\text{d}}$ is the d-state radius that depends on the Wigner-Seitz radius and the linear muffin-tin orbital potential parameters and is related to the d-band center, $l$ is the interatomic distance between an INS and its surrounding TM atoms, $m$ is the mass of an electron with $\hbar^2/m = 7.62$ eVÅ$^2$, and $\eta_{\text{sd}}$ is a universal dimensionless coefficient that is associated with the type of bonding interactions. The characteristic d-state radius $r_{\text{d}}$ of each TM element is obtained from the Solid-State Table in Ref.[37], as shown in Supplementary Table 1.

In the TB model, the $E_{\text{sol}}$ of INSs is proportional to the square of $V_{\text{is}}$[39,42]. In the BCC structure, the second nearest-neighboring (2nn) shell is close to the first nearest-neighboring (1nn) shell with the ratio of interatomic distance $l_{\text{2nn}}/l_{\text{1nn}} = 1.15$. The interatomic coupling of d-d states $V_{\text{dd}}$ is mainly concentrated within one lattice constant, containing the contribution of both the 1nn and 2nn atoms with the ratio of $V_{\text{dd}}^{\text{2nn}}$ to $V_{\text{dd}}^{\text{1nn}}$ as a constant value of $1/2$[37,38]. Meanwhile, Elsässer *et al.* also demonstrated that the interactions between interstitial H and substrate atoms primarily decay within the range of one lattice constant[25]. In the BCC pure metals, the distance between the center of TISs and the 1nn and 2nn atoms is $0.559a$ and $0.901a$, respectively, where $a$ is the lattice constant. The distance between the center of OISs and the octahedral vertices is $0.5a$ and $0.707a$ (both considered as the 1nn atoms), while the distance to the 2nn atoms is $1.118a$. Therefore, the INSs in the TISs of BCC structures is determined by both the 1nn and 2nn atoms (Fig. 1a), while for the OISs the solution is mainly determined by the 1nn atoms (Fig. 1b). The interatomic matrix element $V_{\text{is}}$ decays more slowly with the interatomic distance $l$ than the d-d coupling $V_{\text{dd}}$ between metal atoms ($V_{\text{dd}} \propto r_{\text{d}}^3/l^5$)[37,43], and the ratio of the contributions to $E_{\text{sol}}$ from $V_{\text{is}}^{\text{2nn}}$ to $V_{\text{is}}^{\text{1nn}}$ is approximated as $3/8$. Thus, for the TISs in BCC structures, the general expression of $V_{\text{is}}^2$ between the states of an INS like H and the d-states of TMs is given by:

$$V_{is}^2 = (V_{is}^{1nn})^2 + (V_{is}^{2nn})^2$$

$$= \left(-3.23 \times 7.62 \frac{r_d^{3/2}}{l^{7/2}}\right)_{1nn}^2 + \frac{3}{8}\left(-1.83 \times 7.62 \frac{r_d^{3/2}}{l^{7/2}}\right)_{2nn}^2 \quad (2)$$

where the universal dimensionless coefficient $\eta_{sd}^{1nn}$ and $\eta_{sd}^{2nn}$ are -3.23 and -1.83, respectively[37], and $r_{is}$ is constant for a given INS and thus not considered in this formula for determining the trend of bonding strength in different TMs. For the OISs in BCC structures, the expression of $V_{is}^2$ only contains the contribution of the 1nn atoms, and is $V_{is}^2 = (V_{is}^{1nn})^2 = \left(-3.23 \times 7.62 \frac{r_d^{3/2}}{l^{7/2}}\right)_{1nn}^2$.

For the H adsorption on TM surfaces, the classical d-band center model assumes that the interatomic distance $l$ between H and metal atoms remains constant (which is only reasonable for the late-TMs), indicating that $V_{is}$ is primarily controlled by the $r_d$ of the TM elements[42,44]. In contrast, the interstitial H in RHEAs is influenced by the multiple neighboring constituent elements. The interatomic distance $l$ cannot be simply treated as a constant, and the variation of $l$ also significantly affect $V_{is}$. Therefore, we consider the synergistic effect of $r_d$ and $l$ in $V_{is}$ to present a new physical picture for the interstitial H in RHEAs, which significantly differs from the classical d-band center model that assumes a constant $l$.

Obviously, the interatomic distance $l$ between an interstitial H and neighboring metal atoms in RHEAs is affected by the size of the constituent elements, and is very likely proportional to the atomic radius of TM atoms ($R_{TM}$). Therefore, we use $R_{TM}$ as an approximate substitute of $l$ for each constituent element in RHEAs. As a result, the interatomic matrix element for the s-orbital of H and the d-states of a given constituent element in RHEAs can be expressed as: $V_{is} \propto r_d^{3/2}/l^{7/2} \propto r_d^{3/2}/R_{TM}^{7/2}$. As the introduction of H has a minor effect on the electronic properties of TMs, the interactions between H atoms and TM atoms are assumed predictable using the intrinsic electronic-structure characteristics of TMs, $r_{d,TM}$. For the complex coupling of the interstitial H to multiple constituent elements in RHEAs, we introduce the geometric mean of the $V_{is}^2$ of the neighboring constituent elements to quantify the high-entropy alloying effects on the solution of interstitial H. We thus derive the theoretical expression $V_{is}^2$ for the interstitial solutes like H and their neighboring atoms in RHEAs, based on the classical TB model, as

$$V_{is}^2 = \left(\prod_{i=1}^n V_{is}^{1nn,i}\right)^{2/n} + \left(\prod_{i=1}^n V_{is}^{2nn,i}\right)^{2/n}$$

$$= 3.23 \times 7.62 \left[\prod_{i=1}^n \frac{(r_{d,TM}^i)^{3/2}}{(R_{TM}^i)^{7/2}}\right]_{1nn}^{2/n} + \frac{3}{8} \times 1.83 \times 7.62 \left[\prod_{i=1}^n \frac{(r_{d,TM}^i)^{3/2}}{(R_{TM}^i)^{7/2}}\right]_{2nn}^{2/n} \quad (3)$$

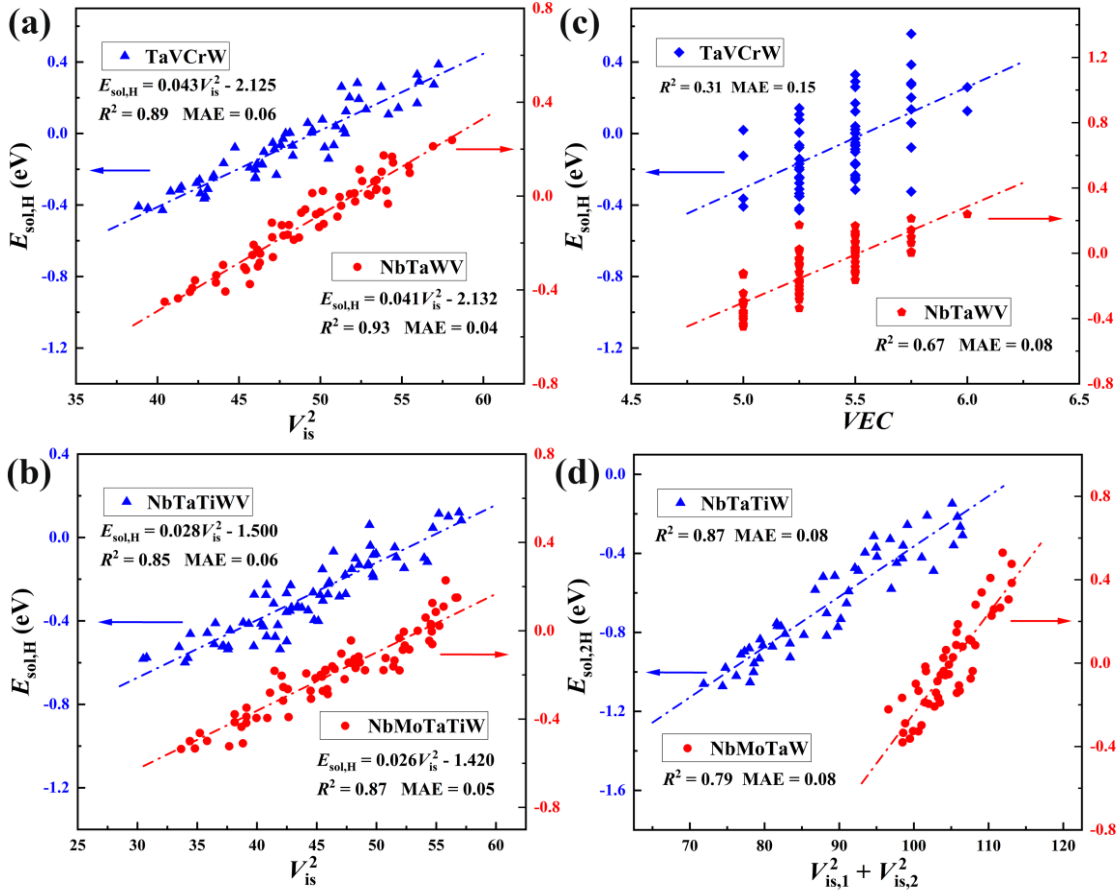

**Fig. 2 The solution energy ($E_{sol}$) of the interstitial H in RHEAs against the interatomic matrix element $V_{is}^2$ and the corresponding *VEC*.** (**a**-**b**) The linear relationship between $E_{sol,H}$ and $V_{is}^2$ in (**a**) TaVCrW and NbTaWV and (**b**) NbTaTiWV and NbMoTaTiW RHEAs. (**c**) The linear relationship between the $E_{sol,H}$ and *VEC* in TaVCrW and NbTaWV. (**d**) The linear relationship between the $E_{sol,2H}$ of two near-neighboring H atoms and the summation of the $V_{is}^2$ of two TISs in NbTaTiW and NbMoTaW RHEAs.

$V_{is}^2$ exhibits a good linear relationship with $E_{sol,H}$ in all calculated RHEAs, including TaVCrW, NbTaWV, NbTiZrW, NbMoTaW, NbTaTiW, NbMoTaTi, NbTaTiWV, NbMoTaTiW, NbMoTaWV and HfNbTaTiW (Fig. 2a, b and Supplementary Fig. 1). This suggests that $V_{is}^2$ effectively captures the complex coupling interactions between interstitial H and the constituent elements in RHEAs. The widely used *VEC* model is limited in certain RHEAs where some constituent elements have the same number of valence electrons[32,33] (Fig. 2c), a challenge that has been solved by $V_{is}^2$. Notably, $V_{is}^2$ can determine the solution energy of interstitial H in the early-TMs with a reasonable accuracy (Supplementary Fig. 2). We also attempt to quantify $E_{sol,H}$ directly using the d-band center, d-band width and d-d coupling $V_{dd}^2$ of the neighboring metal atoms; however, the corresponding accuracy and universality are limited (Supplementary Fig. 3). $V_{is}^2$ is linearly related to the d-band width of metals (Supplementary Fig. 4). More importantly, $V_{is}^2$ further incorporates the s-orbital contribution of H, reflecting the interatomic coupling between INSs and the neighboring TMs, thus generating higher accuracy and universality. For multiple H solution in RHEAs, we examine the case where H atoms occupy two neighboring TISs, with their corresponding atomic configurations illustrated in Supplementary Fig. 5. The results indicate that the total solution energies of two neighboring H atoms ($E_{sol,2H}$) depend linearly on the sum of the individual $E_{sol,H}$, as shown in Supplementary Fig. 6a. Moreover, $E_{sol,2H}$ can be described by the sum of the $V_{is}^2$ of the two neighboring TISs (Fig. 2d and Supplementary Fig. 6b). Therefore, we expect $V_{is}^2$ to be applied to multiple H systems.

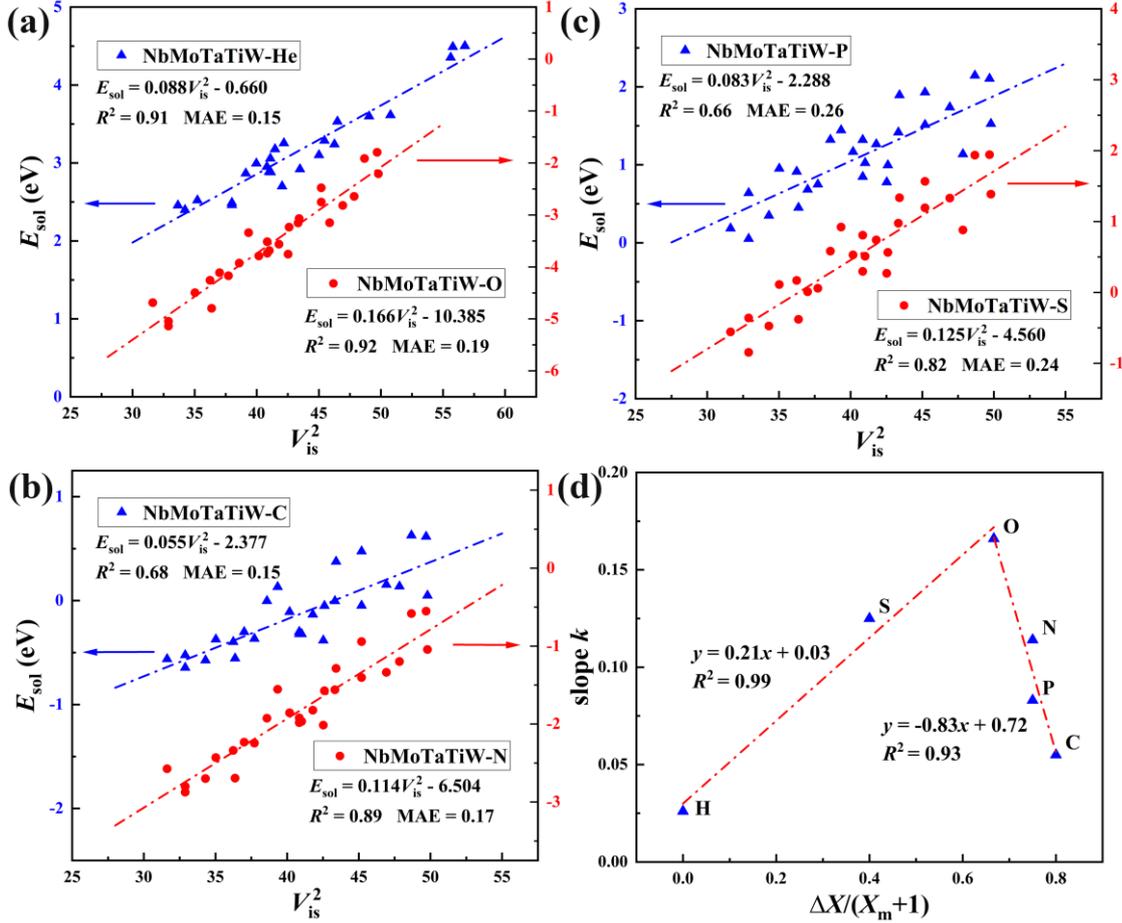

**Fig. 3 Linear correlations between the solution energy ($E_{sol}$) of INSs in NbMoTaTiW and the interatomic matrix element $V_{is}^2$, and the corresponding slope $k$ against the bonding characteristics of INSs.** (**a**-**c**) The linear relationship between the $E_{sol}$ of interstitial (**a**) He and O, (**b**) C and N, and (**c**) P and S with $V_{is}^2$. (**d**) The broken-line relationship between the slope $k$ (that of the linear correlation between $V_{is}^2$ and the $E_{sol}$ of INSs) and the bonding characteristics of different INSs.

$V_{is}^2$ is found to exhibit a wide range of applicability for interstitial He, O, C, N, P, and S atoms in RHEAs. Herein, He atom tends to occupy TISs due to its small size, whereas the larger atomic sizes O, C, N, P, and S atoms prefer OISs. Figure 3 and Supplementary Figure 7 show the linear correlation between the $E_{sol}$ of INSs and $V_{is}^2$. For interstitial He atoms, $V_{is}^2$, which combines the contributions of the 1nn and 2nn atoms, effectively quantifies $E_{sol,He}$ at different TISs in RHEAs (Fig. 3a and Supplementary Fig. 7a, b). For interstitial O, C, N, P, and S atoms, their $E_{sol}$ in RHEAs can be quantified in terms of $V_{is}^2$ between their sp-states and the d-states of the octahedral vertex atoms (both considered as the 1nn atoms) (Fig. 3a-c and Supplementary Fig. 7c-j). These results demonstrate that our assumptions about the similarity between INS-TM interactions and TM-TM interactions are reasonable and accurate, indicating that the TB model is applicable to the INS-TM interactions. Furthermore, we find that the alloying effects of constituent elements in RHEAs on the solution of INSs comply with the mean-field effects, revealing a new physical picture of the coupling mechanism between INSs and RHEAs.

Therefore, the $E_{\text{sol}}$ of INSs in RHEAs can be described by a linear function of $V_{\text{is}}^2$, as follows

$$E_{\text{sol}} = kV_{\text{is}}^2 + b \qquad (4)$$

In this equation, $kV_{\text{is}}^2$ represents the coupling between the states of INSs and the d-states of TMs, with the slope $k$ depending on the type of INSs, while $b$ represents the coupling between the states of INSs and the s-states of TMs, which is a constant for a given INS in a specific RHEA system. Furthermore, we find that for the same RHEAs, the slope $k$ of the linear relationship between $E_{\text{sol}}$ and $V_{\text{is}}^2$ for different INSs is associated with the bonding characteristics of solutes. Based on the first-order approximation of effective medium theory (EMT) and the linear scaling relationship (LSR) for the adsorption energy of H-containing molecules on TM surfaces[45–47], we have developed a concise formula, $\frac{\Delta X}{X_{\text{m}}+1}$, for determining the role of adsorbates in the adsorption energy[48], effectively overcoming the drawback of LSR in understanding the inhomogeneous systems and non-H-containing molecules. In this formula, $\Delta X$ and $X_{\text{m}}$ are the unsaturated bond number and maximum available bond number of the central atom in adsorbates, respectively. Here, we find a clear broken-line relationship between the slope $k$ and $\frac{\Delta X}{X_{\text{m}}+1}$ for different INSs (see Fig. 3d and Supplementary Note 1). These results not only demonstrate the generalization of EMT in understanding the coupling mechanism between INSs and RHEAs, but also provide a tool for clarifying the different behavior of INSs in RHEAs.

Our scheme provides a new perspective for understanding the synergistic effects of multiple alloying elements on the solution of INSs in RHEAs. We find that the interactions of INSs with the alloying elements in RHEAs are determined by both the $r_{\text{d}}$ and $R_{TM}$ of the metal atoms, which differs from the d-band center model of surface adsorption (that assuming a constant adsorption distance on different TMs). Namely, the coupling between INSs and the constituent elements of RHEAs is dominated by the d-band width of the alloying elements and the valence of INSs.

**Interstitial diffusion in RHEAs**

Next, we investigate the diffusion of INSs in NbTaWV, TaVCrW, NbMoTaW, NbTaTiW, NbMoTaTiW and HfNbTaTiW RHEAs. The driving force ($E_{\text{d}}$) of H diffusion is determined by the $E_{\text{sol,H}}$ difference between the IS and FS sites. We find that $E_{\text{d}}$ can be quantified by the difference between the $V_{\text{is}}^2$ at the IS and FS sites, defined as $\Delta V_{\text{is}}^2 = V_{\text{is,FS}}^2 - V_{\text{is,IS}}^2$. Clearly, $\Delta V_{\text{is}}^2$ exhibits a strong predictive capability to $E_{\text{d}}$ along both the T-T and T-O-T paths (Fig. 4a, b and Supplementary Fig. 8), showing that a greater difference in $V_{\text{is}}^2$ between TISs corresponds to a stronger tendency for H diffusion in RHEAs. Furthermore, $\Delta V_{\text{is}}^2$ is also applicable to the $E_{\text{d}}$ of interstitial O in RHEAs (Supplementary Fig. 12a).

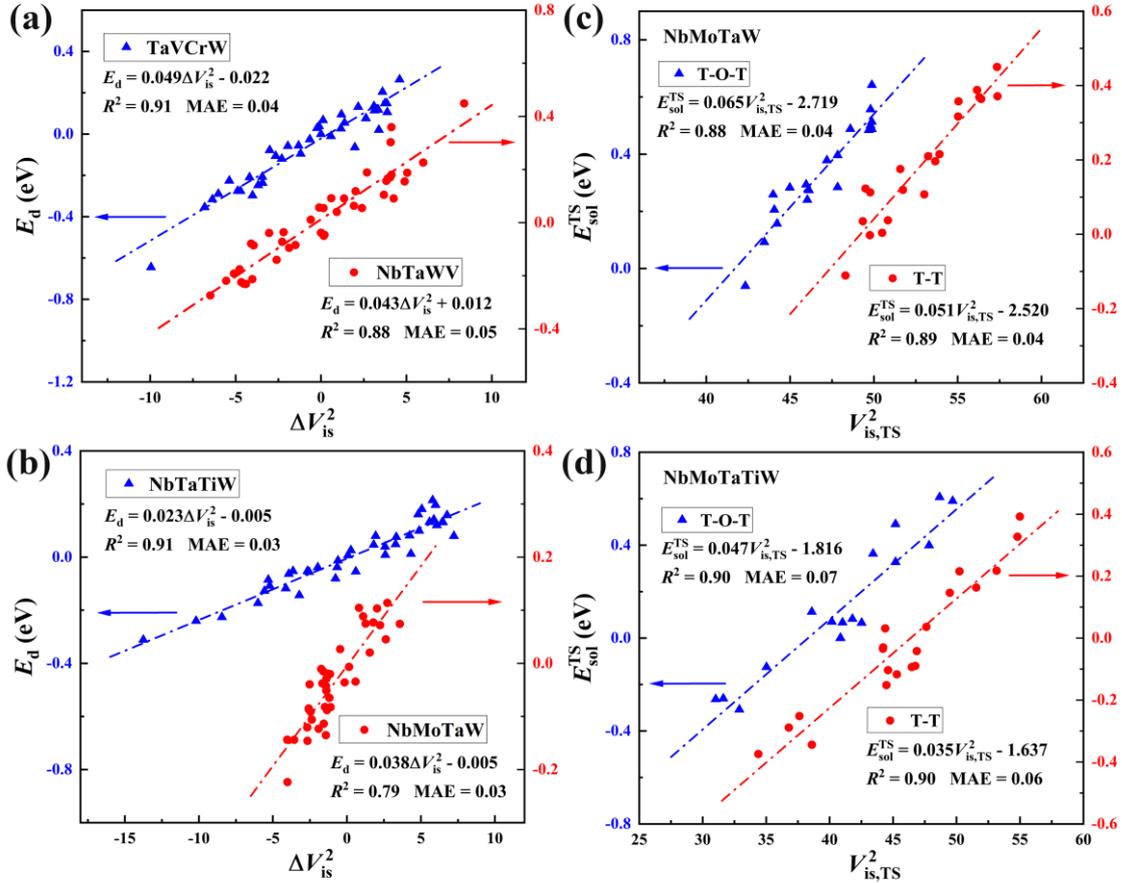

**Fig. 4 The H diffusion driving force ($E_d$) and the solution energy at transition-state sites ($E_{sol}^{TS}$) in RHEAs.** (**a-b**) $E_d$ against the difference of $V_{is}^2$ between the final-state and initial-state sites ($\Delta V_{is}^2$) in (**a**) TaVCrW and NbTaWV, (**b**) NbTaTiW and NbMoTaW RHEAs. (**c-d**) $E_{sol}^{TS}$ as a function of $V_{is,TS}^2$ in (**c**) NbMoTaW and (**d**) NbMoTaTiW RHEAs.

For the diffusion barrier $E_b$ of H in RHEAs, we observe that for a given initial TIS, $E_b$ along the T-T path is consistently lower than that of the T-O-T path (Supplementary Fig. 9). This suggests that H diffusion occurs predominantly along the <110> direction in RHEAs, which is consistent with the behavior in BCC TMs[23,26]. As shown in Supplementary Fig. 10, the transition-state (TS) structure features a coplanar double tetrahedral configuration with five vertices for the T-T path, but an octahedral configuration with six vertices for the T-O-T path. For the TS site along the T-T path, the distance between interstitial H and its 2nn atoms ranges from $0.90a$ to $1.05a$, which is approximately within one lattice constant. Thus, for the TS site along the T-T path, $E_{sol}^{TS}$ is determined by both the 1nn and 2nn atoms, while for the T-O-T path, $E_{sol}^{TS}$ is determined by the 1nn atoms. As shown in Fig. 4c, d and Supplementary Fig. 11, $V_{is,TS}^2$ effectively quantifies the $E_{sol}^{TS}$ at the TS sites for both T-O-T and T-T paths in RHEAs, suggesting that $V_{is}^2$ can capture the interactions of interstitial H with the alloying atoms at the RHEAs interstitial sites with different geometric structures. For a given initial TIS with interstitial H, among the candidate diffusion paths, the TS site with the smallest $V_{is,TS}^2$ value indicates the lowest $E_{sol}^{TS}$, which corresponds to the lowest $E_b$. Therefore, the H diffusion route from a given initial site can be determined by the distribution of the surrounding atoms based on $V_{is,TS}^2$. Furthermore, $V_{is,TS}^2$ is also applicable to the diffusion of interstitial O in RHEAs (Supplementary Fig. 12b). Namely, our approach allows for the continuous updates, enabling an ongoing prediction of the overall diffusion route.

## Discussion

We try to build a picture for the coupling of INSs and their surrounding metal atoms in RHEAs based on the News-Anderson-Grimley (NAG) model[42,49–51]. The NAG model is analytical and greatly helpful in understanding adsorbate-surface bonding because of its conceptual simplicity. It shows that the valence bandwidth of TMs determines the bonding strength for a given molecule on TMs. Figure 5a, b illustrates the coupling rule of INSs and metal atoms, by approximating the metal's density of states (DOS) with a semi-ellipse. The bonding function of INSs and metal atoms is in the form of $(\varepsilon - \varepsilon_{is})/V^2$ (blue dotted line), where $\varepsilon$ and $\varepsilon_{is}$ are the energy of the metal and solute electron state, and $V$ is the coupling matrix with $V^2$ corresponding to the coupling strength. With the Hilbert transform of the d-band DOS (red curve), one can deduce the solute state splitting or broadening by counting the crossing number of bonding function and the Hilbert transform[49,51]. The broadening occurs if the crossing number is once and the splitting occurs if the crossing number is more than once.

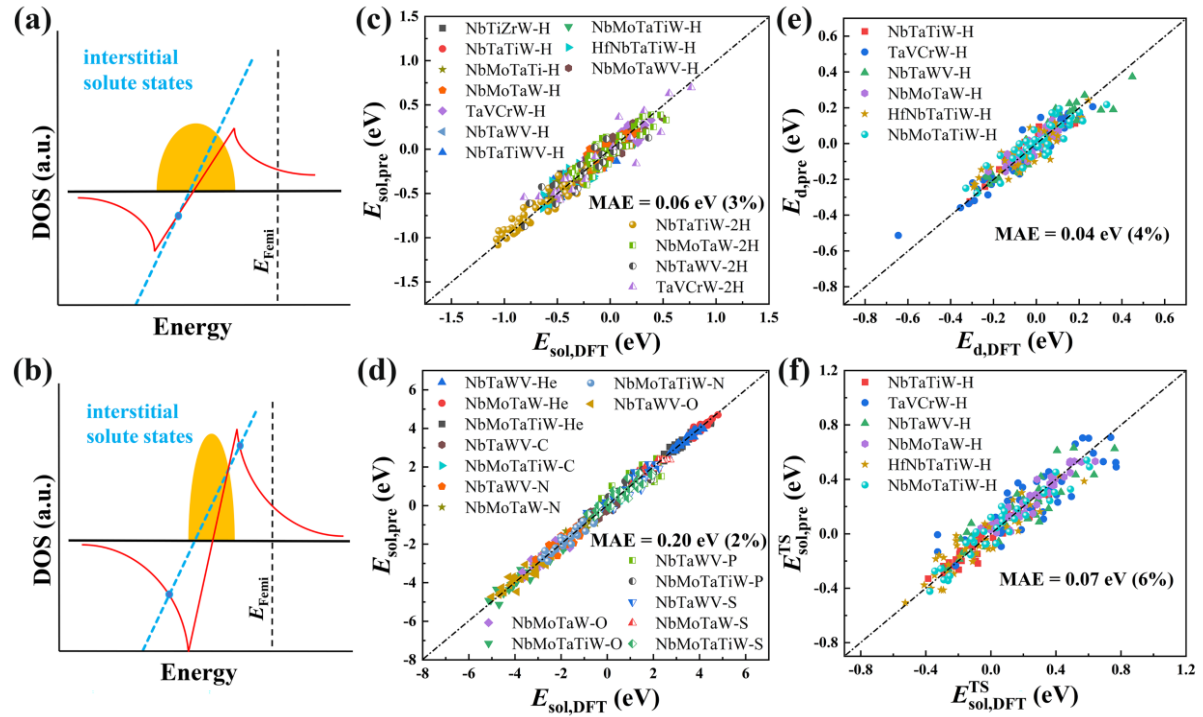

**Fig. 5 The physical picture of $V_{is}^2$ and its prediction accuracy.** (a-b). Schematic diagram of the News-Anderson-Grimley model for the case of the broad metal band in (a) and the case of the narrow metal band in (b). The yellow semi-ellipses represent the density of states

(DOS) distribution of metals, the red curve illustrates the Hilbert transform of the DOS, and the blue dotted lines represent the bonding function between the solute and metal. (**c-f**) Comparison of the predicted results with the DFT-calculated ones for (**c**, **d**) interstitial solution energy ($E_{sol}$), (**e**) diffusion driving force ($E_d$) and (**f**) transition-state solution energy ($E_{sol}^{TS}$) in various RHEAs with the DFT-calculated results.

The slope of the bonding function $(\varepsilon - \varepsilon_{is})/V^2$ is determined by $1/V^2$, while the characteristic slope of Hilbert transform (within the yellow semi-ellipses) is determined by the bandwidth $W$ in the form of $1/W$. If $W > V^2$, the bonding function can only intersect with the Hilbert transform once, generating a broadened rehybridized solute state. If $W < V^2$, the bonding function intersects with the Hilbert transform more than once, resulting in the solute state splitting that generates both bonding and antibonding rehybridized states (marked Fig. 5b), with the bonding-antibonding gap $\Delta E \propto V^2/W$. For a given solute bonding with the BCC metal atoms (that generally have $\varepsilon_d > \varepsilon_{is}$), the decrease of bandwidth $W$ enlarges $\Delta E$ by lowering the energy of bonding states and enhancing the energy of anti-bonding states. Relative to the Fermi level, the lower bonding states strengthen the bonding, and the higher anti-bonding states corresponding to less filled states also strengthen the bonding. Therefore, the decrease of bandwidth $W$ results in an increased bonding strength between INSs and metals, as we found in RHEAs. Obviously, the NAG model can be well generalized into the coupling of INSs and metal atoms in RHEAs, explaining our findings about the dominant role of d-band width $W_d$ in determining the $E_{sol}$ of INSs in RHEAs.

Notably, our model can effectively predict the $E_{sol}$ of interstitial H, He, O, C, N, P, and S in ten RHEAs, as well as the $E_d$ and $E_{sol}^{TS}$ of interstitial H and O, demonstrating the broad applicability and strong predictive accuracy. As shown in Fig. 5c, $V_{is}^2$ achieves a site-to-site prediction of $E_{sol}$ with an overall mean absolute error (MAE) of 0.06 eV (3%) for interstitial H, which is nearly identical to the MAE of 0.068 eV reported by the previous machine-learning model[36]. Additionally, $V_{is}^2$ has an MAE of 0.20 eV (2%) in determining the $E_{sol}$ of interstitial He, O, C, N, P, and S across various RHEA systems (Fig. 5d), displaying a similar accuracy to the CE model[34]. However, the parameters involved in the CE method generally lack clear physical significance. In contrast, our descriptor $V_{is}^2$, derived from TB models and reflecting the d-band width of metals, demonstrates broad applicability across different systems with a solid physical picture. For the diffusion of INSs, our model provides a quantitative prediction method for $E_d$ and $E_{sol}^{TS}$, achieving an overall MAE of 0.04 eV (4%) and 0.07 eV (6%) for interstitial H (Fig. 5e, f), and 0.17 eV (7%) and 0.21 eV (7%) for interstitial O (Supplementary Fig. 12c, d), respectively, thereby enabling the continuous prediction of INSs diffusion routes.

Our model not only effectively quantifies the synergistic effects of multiple elements of RHEAs on the solution and diffusion of INSs, but also effectively distinguishes the contributions of each constituent element in RHEAs. Our model indicates that the elements with the smaller $V_{is}^2$ are more favorable for the solution of INSs in RHEAs (see Supplementary Table 1 for $V_{is}^2$). Previous studies found that in the Reverse Monte Carlo structure of TiVNbD$_{5.7}$, deuterium (D) occupies the highest proportion in TISs with four Ti atoms, along with a higher proportion in TISs with four V atoms compared to those with four Nb atoms[33]. These findings are consistent with the prediction of our model, as the order of $V_{is}^2$ values in TiVNb is $V_{is}^2(Ti) < V_{is}^2(V) < V_{is}^2(Ta)$. Meanwhile, the thermodynamic model developed by Zepon shown that the H solution stability in TiVZrNbHf is higher than that in TiZrNbHfTa based on the plateau enthalpy ($\Delta H_{\text{plat}}$) of H sorption[52], suggesting that V element is more favorable for H solution than Ta element, which is in agreement with the prediction of our model. Moreover, our model captures the effect of element concentration on the stability of alloy hydrides. In TiZrHfMoNb, the Mo element exhibit the highest $V_{\text{is}}^2$ value. Thus, an increase in Mo concentration reduces the stability of interstitial H in TiZrHfMoNb, which corresponds to the experimental findings that the thermal stability of TiZrHfMoNb hydride decreases with increasing Mo content[53]. More importantly, our model can deduce the effect of INSs on the stability of RHEAs observed in experiments. Lu *et al.* found that 2.0 at% O doping in TiZrHfNb leads to unprecedented improvements in both strength and ductility of HEAs[54]. Our model indicates that the elements with the smaller $V_{\text{is}}^2$ are more favorable for O solution in RHEAs. In TiZrHfNb, the Ti and Zr elements, which have relatively smaller $V_{\text{is}}^2$ values, facilitate the solution and accumulation of O atoms, leading to the formation of (O, Ti, Zr)-rich complexes upon the incorporation of 2.0 at% interstitial O, as the experiments found. Moreover, our model elucidates the physical mechanism by which trace interstitial O and N promote the formation of α-Ti precipitates in TiVNbTa during high temperature aging. In TiVNbTa, the Ti element exhibits the smallest $V_{\text{is}}^2$ value, making its vicinity more favorable for the solution of interstitial O and N. The accumulation of these interstitial impurities in the Ti-rich region induces the formation of new phases, corresponding to the experimental observations that interstitial O and N in TiVNbTa can form α-Ti precipitates[55]. These results imply that our model offers new insights into the formation of precipitate phases induced by the accumulation of INSs.

In summary, we build an analytic model to determine the stability and diffusivity of INSs in RHEAs, which derives the interatomic matrix element $V_{\text{is}}^2$ between INSs and constituent elements in the framework of the tight-binding models, by accounting for the bonding length between INSs and their neighboring atoms using the atomic radius of the neighbors in elemental states. This model yields a linear correlation between $V_{\text{is}}^2$ and the energetics of INSs in RHEAs, with the prefactors determined by the valence properties of INSs, revealing that the coupling of INSs and constituent elements depends on the d-band width of INSs' neighboring metal atoms and the bonding properties of INSs in free states and interstitial states. Our scheme is a development of the classical tight-binding model and effective medium theory, explicitly clarifying the similarity and difference for the bonding properties of the INSs at bulk RHEAs and metal surfaces. This model is predictive and descriptive for the theoretical site-to-site energetics of interstitial H, He, O, C, N, P, and S atoms in ten different RHEAs and a series of experimental observations, which provides new perspectives for understanding the INSs' behavior in RHEAs and solid design principles for developing high-performance RHEAs.

## Methods

In this work, all the density functional theory (DFT) calculations were performed by Vienna Ab initio Simulation Package (VASP) code[56] with the projector augmented wave (PAW) approach[57] and Perde−Burke−Ernzerhof (PBE) functional[58]. The lattice constants of RHEAs with different element compositions were calculated using Vegard's law[59]. The 4×4×4 supercells containing 128 atoms for equimolar quaternary RHEAs (NbMoTaW, NbTaTiW, NbMoTaTi, NbTaWV, NbTiZrW and TaVCrW) and 4×4×5 supercells containing 160 atoms for equimolar quinary RHEAs (NbMoTaWV, NbTaTiWV, NbMoTaTiW and HfNbTaTiW) were constructed using the special quasi-random structure (SQS) method[60,61]. In our calculations, the extensive tests allow us to adopt plane-wave cutoff energy of 400 eV for interstitial H systems, 450 eV for interstitial O, C, N, P, and S systems, and 500 eV for interstitial He systems (test shown in Supplementary Fig. 13), as well as a 3×3×3 k-point mesh for sampling the Brillouin using the Monkhorst-Pack method[62]. All atomic positions were fully relaxed with a fixed lattice volume during geometry optimization (see the comparison with lattice optimization in Supplementary Fig. 14a, b). The energies and atomic forces of all calculations were converged within $10^{-5}$ eV and 0.01 eV/Å. We also use a smearing width of 0.1 eV for the Fermi smearing function to facilitate the convergence. The zero-point corrections were not included due to their approximately unchanged effects (see the comparison with zero-point energy corrections in Supplementary Fig. 14c, d). The climbing image nudging elastic band (CI-NEB) method[63] was used to calculate the diffusion energy of INSs in NbMoTaW, NbTaTiW, NbTaWV, TaVCrW, NbMoTaWV and HfNbTaTiW (Supplementary Fig. 14e).

The interstitial solution energy ($E_{sol}$) is calculated using the following formula

$$E_{sol} = E_{HEA+I} - E_{HEA} - E_I \tag{5}$$

where $E_{HEA+I}$ is the total energy of the relaxed HEAs embedding INSs, $E_{HEA}$ is the energy of the relaxed HEA, and $E_I$ is the energy of INSs.

## Data availability

All data are available from the corresponding author upon reasonable request.

## References

1. Yuan, X. *et al.* Effects of trace elements on mechanical properties of the TiZrHfNb high-entropy alloy. *J. Mater. Sci. Technol.* **152**, 135–147 (2023).
2. Song, J. & Curtin, W. A. Atomic mechanism and prediction of hydrogen embrittlement in iron. *Nat. Mater.* **12**, 145–151 (2013).
3. Huang, Z., Chen, F., Shen, Q., Zhang, L. & Rupert, T. J. Uncovering the influence of common nonmetallic impurities on the stability and strength of a Σ5 (310) grain boundary in Cu. *Acta Mater.* **148**, 110–122 (2018).
4. Wu, C. *et al.* Hydrogen-assisted spinodal decomposition in a TiNbZrHfTa complex concentrated alloy. *Acta Mater.* **285**, 120707 (2025).
5. Kwak, N. *et al.* Bimodal structured chromium-tungsten composite as plasma-facing materials: Sinterability, mechanical properties, and deuterium retention assessment. *Acta Mater.* **262**, 119453 (2024).
6. Zhou, X.-Y. *et al.* Formation and strengthening mechanism of ordered interstitial complexes in multi-principle element alloys. *Acta Mater.* **281**, 120364 (2024).
7. Casillas-Trujillo, L. *et al.* Interstitial carbon in bcc HfNbTiVZr high-entropy alloy from

first principles. *Phys. Rev. Mater.* **4**, 123601 (2020).
8. Zhang, B., An, Y., Liu, C., Ding, J. & Ma, E. Tetrahedral versus octahedral interstitial sites for oxygen solutes in NbTiZr medium-entropy alloy. *J. Mater. Sci. Technol.* **241**, 229–237 (2026).
9. Zhang, Q. *et al.* Radiation-enhanced precipitation and the impact on He bubble formation in V-Ti-based refractory alloys containing interstitial impurities. *J. Nucl. Mater.* **596**, 155078 (2024).
10. Kong, L., Cheng, B., Wan, D. & Xue, Y. A review on BCC-structured high-entropy alloys for hydrogen storage. *Front. Mater.* **10**, 1135864 (2023).
11. Li, X. *et al.* Hydrogen embrittlement and failure mechanisms of multi-principal element alloys: A review. *J. Mater. Sci. Technol.* **122**, 20–32 (2022).
12. Marques, F., Balcerzak, M., Winkelmann, F., Zepon, G. & Felderhoff, M. Review and outlook on high-entropy alloys for hydrogen storage. *Energy Environ. Sci.* **14**, 5191–5227 (2021).
13. Senkov, O. N., Wilks, G. B., Scott, J. M. & Miracle, D. B. Mechanical properties of Nb25Mo25Ta25W25 and V20Nb20Mo20Ta20W20 refractory high entropy alloys. *Intermetallics* **19**, 698–706 (2011).
14. Senkov, O. N., Wilks, G. B., Miracle, D. B., Chuang, C. P. & Liaw, P. K. Refractory high-entropy alloys. *Intermetallics* **18**, 1758–1765 (2010).
15. Senkov, O. N., Gorsse, S. & Miracle, D. B. High temperature strength of refractory complex concentrated alloys. *Acta Mater.* **175**, 394–405 (2019).
16. Zhang, Y. *et al.* Microstructures and properties of high-entropy alloys. *Prog. Mater. Sci.* **61**, 1–93 (2014).
17. George, E. P., Raabe, D. & Ritchie, R. O. High-entropy alloys. *Nat. Rev. Mater.* **4**, 515–534 (2019).
18. Miracle, D. B. & Senkov, O. N. A critical review of high entropy alloys and related concepts. *Acta Mater.* **122**, 448–511 (2017).
19. Halpren, E., Yao, X., Chen, Z. W. & Singh, C. V. Machine learning assisted design of BCC high entropy alloys for room temperature hydrogen storage. *Acta Mater.* **270**, 119841 (2024).
20. You, Y.-W. *et al.* Dissolving, trapping and detrapping mechanisms of hydrogen in bcc and fcc transition metals. *AIP Adv.* **3**, 012118 (2013).
21. Wipf, H. Solubility and diffusion of hydrogen in pure metals and alloys. *Phys. Scr.* **T94**, 43 (2001).
22. Liu, W.-G. *et al.* Theoretical study of the interaction between hydrogen and 4d alloying atom in nickel. *Nucl. Sci. Tech.* **28**, 82 (2017).
23. Liu, Y.-L., Zhang, Y., Luo, G.-N. & Lu, G.-H. Structure, stability and diffusion of hydrogen in tungsten: A first-principles study. *J. Nucl. Mater.* **390–391**, 1032–1034 (2009).
24. Kong, X.-S. *et al.* First-principles calculations of hydrogen solution and diffusion in tungsten: Temperature and defect-trapping effects. *Acta Mater.* **84**, 426–435 (2015).
25. Elsässer, C., Fähnle, M., Schimmele, L., Chan, C. T. & Ho, K. M. Range of forces on host-metal atoms around interstitial hydrogen in Pd and Nb. *Phys. Rev. B* **50**, 5155–5159 (1994).

26. Qin, J. *et al.* Dissolution, diffusion, and penetration of H in the group VB metals investigated by first-principles method. *Int. J. Hydrogen Energy* **44**, 29083–29091 (2019).
27. Ohsawa, K., Eguchi, K., Watanabe, H., Yamaguchi, M. & Yagi, M. Configuration and binding energy of multiple hydrogen atoms trapped in monovacancy in bcc transition metals. *Phys. Rev. B* **85**, 094102 (2012).
28. Moore, C. M. *et al.* Hydrogen accommodation in the TiZrNbHfTa high entropy alloy. *Acta Mater.* **229**, 117832 (2022).
29. Yang, T.-R. *et al.* Exploring the inhibitory effect of WTaVCr high-entropy alloys on hydrogen retention: From dissolution, diffusion to desorption. *J. Nucl. Mater.* **601**, 155346 (2024).
30. Ren, X. L. *et al.* Hydrogen solution in high-entropy alloys. *Phys. Chem. Chem. Phys.* **23**, 27185–27194 (2021).
31. Ek, G. *et al.* Elucidating the effects of the composition on hydrogen sorption in TiVZrNbHf-based high-entropy alloys. *Inorg. Chem.* **60**, 1124–1132 (2021).
32. Nygård, M. M. *et al.* Counting electrons - A new approach to tailor the hydrogen sorption properties of high-entropy alloys. *Acta Mater.* **175**, 121–129 (2019).
33. Nygård, M. M. *et al.* Local order in high-entropy alloys and associated deuterides – a total scattering and Reverse Monte Carlo study. *Acta Mater.* **199**, 504–513 (2020).
34. Borges, P. P. P. O., Ritchie, R. O. & Asta, M. Chemical trends favoring interstitial cluster formation in bcc high-entropy alloys from first-principles calculations. *Acta Mater.* **294**, 121091 (2025).
35. Shuang, F. *et al.* Decoding the hidden dynamics of super-Arrhenius hydrogen diffusion in multi-principal element alloys via machine learning. *Acta Mater.* **289**, 120924 (2025).
36. Korostelev, V., Wagner, J. & Klyukin, K. Simple local environment descriptors for accurate prediction of hydrogen absorption and migration in metal alloys. *J. Mater. Chem. A* **11**, 23576–23588 (2023).
37. Harrison, W. A. *Electronic Structure and the Properties of Solids: The Physics of the Chemical Bond*. (Dover Publications, Inc, New York, 1989).
38. Desjonquères, M.-C. & Spanjaard, D. *Concepts in Surface Physics*. vol. 30 (Springer Berlin Heidelberg, 1993).
39. Hammer, B. & Nørskov, J. K. Theory of Adsorption and Surface Reactions. in *Chemisorption and Reactivity on Supported Clusters and Thin Films* (eds. Lambert, R. M. & Pacchioni, G.) 285–351 (Springer Netherlands, Dordrecht, 1997).
40. Wu, C. *et al.* Hydrogen accommodation and its role in lattice symmetry in a TiNbZr medium-entropy alloy. *Acta Mater.* **288**, 120852 (2025).
41. Kitte, C. *Introduction to Solid State Physics*. (John Wiley & Sons, Inc, Hoboken, NJ, 2005).
42. Vojvodic, A., Nørskov, J. K. & Abild-Pedersen, F. Electronic structure effects in transition metal surface chemistry. *Top. Catal.* **57**, 25–32 (2014).
43. Harrison, W. A. & Froyen, S. Universal linear-combination-of-atomic-orbitals parameters for d -state solids. *Phys. Rev. B* **21**, 3214–3221 (1980).
44. Hammer, B. & Nørskov, J. K. Why gold is the noblest of all the metals. *Nature* **376**, 238–240 (1995).
45. Nørskov, J. K. & Lang, N. D. Effective-medium theory of chemical binding: Application

to chemisorption. *Phys. Rev. B* **21**, 2131–2136 (1980).
46. Nørskov, J. K. Covalent effects in the effective-medium theory of chemical binding: Hydrogen heats of solution in the 3 d metals. *Phys. Rev. B* **26**, 2875–2885 (1982).
47. Abild-Pedersen, F. *et al.* Scaling properties of adsorption energies for hydrogen-containing molecules on transition-metal surfaces. *Phys. Rev. Lett.* **99**, 016105 (2007).
48. Gao, W. *et al.* Determining the adsorption energies of small molecules with the intrinsic properties of adsorbates and substrates. *Nat. Commun.* **11**, 1196 (2020).
49. Newns, D. M. Self-consistent model of hydrogen chemisorption. *Phys. Rev.* **178**, 1123–1135 (1969).
50. Anderson, P. W. Localized magnetic states in metals. *Phys. Rev.* **124**, 41–53 (1961).
51. Grimley, T. B. The indirect interaction between atoms or molecules adsorbed on metals. *Proc. Phys. Soc.* **90**, 751–764 (1967).
52. Zepon, G., Silva, B. H., Zlotea, C., Botta, W. J. & Champion, Y. Thermodynamic modelling of hydrogen-multicomponent alloy systems: Calculating pressure-composition-temperature diagrams. *Acta Mater.* **215**, 117070 (2021).
53. Shen, H. *et al.* Compositional dependence of hydrogenation performance of Ti-Zr-Hf-Mo-Nb high-entropy alloys for hydrogen/tritium storage. *J. Mater. Sci. Technol.* **55**, 116–125 (2020).
54. Lei, Z. *et al.* Enhanced strength and ductility in a high-entropy alloy via ordered oxygen complexes. *Nature* **563**, 546–550 (2018).
55. Liu, J. *et al.* Origin of age softening in the refractory high-entropy alloys. *Sci. Adv.* **9**, eadj1511 (2023).
56. Kresse, G. & Furthmüller, J. Efficient iterative schemes for *ab initio* total-energy calculations using a plane-wave basis set. *Phys. Rev. B* **54**, 11169–11186 (1996).
57. Blöchl, P. E. Projector augmented-wave method. *Phys. Rev. B* **50**, 17953–17979 (1994).
58. Perdew, J. P., Burke, K. & Ernzerhof, M. Generalized gradient approximation made simple. *Phys. Rev. Lett.* **77**, 3865–3868 (1996).
59. Denton, A. R. & Ashcroft, N. W. Vegard's law. *Phys. Rev. A* **43**, 3161–3164 (1991).
60. Zunger, A., Wei, S.-H., Ferreira, L. G. & Bernard, J. E. Special quasirandom structures. *Phys. Rev. Lett.* **65**, 353–356 (1990).
61. Van De Walle, A. *et al.* Efficient stochastic generation of special quasirandom structures. *Calphad* **42**, 13–18 (2013).
62. Chadi, D. J. Special points for Brillouin-zone integrations. *Phys. Rev. B* **16**, 1746–1747 (1977).
63. Henkelman, G., Uberuaga, B. P. & Jónsson, H. A climbing image nudged elastic band method for finding saddle points and minimum energy paths. *J. Chem. Phys.* **113**, 9901–9904 (2000).

## Acknowledgments

The authors are thankful for the support from the National Natural Science Foundation of China (Nos. 22173034, 11974128, 52130101), the Opening Project of State Key Laboratory of High-Performance Ceramics and Superfine Microstructure (SKL202206SIC), the Program

## Author contributions

Wang Gao and Qing Jiang conceived the original idea and designed the strategy. Wang Gao derived the models with the contribution of Qianxi Zhu. Wang Gao and Qianxi Zhu analyzed the results and wrote the manuscript together. Qianxi Zhu performed the calculations, drew figures and prepared the Supplementary Information with the contribution of Wang Gao. Qianxi Zhu, Wang Gao and Qing Jiang have discussed and approved the results and conclusions of this article.

## Competing interests

The authors declare no competing interests.